\documentclass[aps]{JHEP3}

\newcommand{\ep}{\epsilon}

\newcommand{\vphi}{\varphi}

\newcommand{\pa}{\partial}
\newcommand{\td}{\tilde}

\newcommand{\Sla}[1]{\slash\!\!\!\! #1}
\newcommand{\beq}[1]{\begin{eqnarray}\label{#1}}
\newcommand{\eeq}{\end{eqnarray}}
\newcommand{\net}{{\cal N}=2}

\title{Gauge/Gravity Duality, Green Functions of ${\cal N}=2$ SYM
and Radial/Energy-Scale Relation}

\author{Xiao-Jun Wang\\
Institute of Theoretical Physics, Beijing, 100080, P.R.China \\
E-mail: wangxj@itp.ac.cn}

\author{Seng Hu\\Department of Mathematics, University of Science and
Technology of China, \\
AnHui, HeFei, 230026, P.R. China\\
E-mail: shu@ustc.edu.cn}

\abstract{We consider supergravity configuration of D5 branes
wrapped on supersymmetric 2-cycles and use it to calculate
one-point and two-point Green functions of some special operators
in $\net$ super Yang-Mills theory. We show that Green functions
obtained from supergravity include two very different parts. One
of them corresponds to perturbative results of quantum field
theory, and another is a non-perturbative effect which
corresponds to contribution from instantons with fractional
charge. Comparing Green functions obtained from supergravity and
gauge theory, we obtain radial/energy-scale relation for this
gauge/gravity correspondence with $\net$ supersymmetry. This
relation leads right $\beta$-function of $\net$ SYM from
supergravity configuration.}


\begin{document}

\section{Introduction}

Recently, Maldacena's conjecture on $AdS_5$/CFT$_4$
correspondence\cite{Mald98} has been thoroughly investigated in
large $N$ limit, and has been shown that it is a precise duality
between the ${\cal N}=4$ super Yang-Mills (SYM) theory in
4-dimension and the Type IIB superstring in $AdS_5\times S^5$. It
makes more and more physicists believe that a lot of information
of gauge theories can be obtained by studying their dual gravity
background produced by stacks of D-branes even though
supersymmetry and/or conformal invariance are (partly) broken.

The simplest extension of $AdS_5$/CFT$_4$ correspondence is to
consider $\net$ SYM and its supergravity (SUGRA) duals. This dual
SUGRA configurations have been studied in many literatures.
Usually, there are two ways to reduce the number of preserved
supercharges. One of them is to place a stack of D3 branes at the
apex of an orbifold\cite{Orbifold} or of a
conifold\cite{Conifold}. Both for the orbifold and the conifold
the conformal invariance can be naturally broken by means of
fractional D3 branes\cite{Fbrane}. Then one can realize a $\net$
non-conformal SYM theory in four dimensions. The recent reviews
on this approach can be found in Refs.\cite{BHreview}. Another
method is to consider D-branes whose world-volume is partially
wrapped on a supersymmetric cycle inside a K3 or a Calabi-Yau
manifold. The unwrapped part of the brane world-volume remains
flat and supports a gauge theory. Meanwhile, the normal bundle to
the wrapped D-branes has to be partially twisted\cite{BVS96} in
order to make some world-volume fields become massive and
decouple. This method has been first used in Ref.\cite{MN01} to
study pure ${\cal N}=1$ SYM theory in four dimensions, and later
it has been generalized to study $\net$ SYM in four
dimension\cite{GKM01,BCZ01} and other cases with different
space-time dimesions and different numbers of
supersymmetry\cite{review2}. The purpose of this paper is to
study some Green functions of dimension four operator\footnote{In
this paper we focus on a non-conformal field theory. So that the
dimension of operators means that mass dimension rather than
conformal dimension} of $\net$ SYM by means of wrapped D5 brane
configuration, and to find radial/energy-scale relation of this
gauge/gravity duality.

A crucial ingredient for the gauge/gravity correspondence is the
relation between the radial parameter of the SUGRA solution and
the energy scales of the gauge theory. In particular, authors of
Ref.\cite{BVV00} have established a formal relation between
radial flow of 5-d gravity and renormalization running of 4-d
gauge theory. However, in general it maybe ambigous and difficult
to establish a exact radial/energy-scale relation for
non-conformal theories\cite{PP98}. Recently, for pure $\net$ and
${\cal N}=1$ SYM this difficulty has been partly overcome due to
the work of Di Vecchia, Lerda and Merlatti (DLM)\cite{DLM02}. They
studied $\beta$-function and chiral anomaly of pure $\net$
SYM\footnote{The paper of DLM also includes study on
$\beta$-function and chiral anomaly of ${\cal N}=1$ SYM. However,
in this paper we only focus on $\net$ SYM due to the reason to be
mentioned in last section.} via studying their dual SUGRA
configurations which is constructed by D5 branes wrapped a
supersymmetry 2-cycle. According to some naive arguments, they
also established a simple radial/energy-scale relation for this
gauge/gravity correspondence configuration. Their result should be
right, perturbatively at least, since it leads to right
$\beta$-function of $\net$ SYM which can be obtained perturbative
calculation of quantum field theory (QFT). However, It is still
necessary to study this relation in different way. This is just
one of main aims of this present paper.

Our idea is motivated by UV/IR relation\cite{PP98} in
gauge/gravity duality. Concretely, the near boundary limit (in
most case, it is large radial limit), i.e., IR limit of SUGRA
description corresponds to UV limit of dual SYM. This relation
will explicitly exhibit in correlation functions (or Green
function for non-conformal) resulted from SUGRA description and
QFT method. In other words, if we calculate some Green functions
of SYM in terms of perturbative QFT method, it will in general
involve an UV cut-off. Meanwhile, when we calculate these Green
functions via dual SUGRA description, it will in general involve
an large radial cut-off. Therefore, we can investigate divergence
behavior of Green functions resulted from SUGRA and QFT
respectively, and compare these terms possessing same pole
behavior. Then we will obtain an exact radial/energy-scale
relation for this gauge/gravity correspondence configuration.

Since $\net$ SYM is a non-conformal theory, its Green functions
obtained from SUGRA description must appear some features that are
very different from ${\cal N}=4$ SYM. A direct feature is that
they should include many non-perturbative effects because dual
SUGRA describes a strong coupled $\net$ SYM in fact. At QFT side,
it has been well-known that these non-perturbative effects should
be induced by instanton\cite{Seiberg88}, but they can not be
obtained via perturbative calculation of QFT method. In this
paper, we will see that this effect can indeed be yielded from
dual SUGRA configuration. In particular, the one-point function of
dimension four operator obtained from SUGRA implies that gauge
bosons and scalars will condensate, and it is instanton effect
entirely.

The paper is organized as follows. In section 2 we review the
$\net$ SUGRA solution\cite{GKM01,BCZ01,DLM02} in a first order
formalism. It creates a background of D5 branes wrapped on
supersymmetry 2-cycle. In section 3 we consider the fluctuations
on this background, and find the solution of these fluctuation
field whose boundary value couples to dimension four operator of
$\net$ SYM. In section 4, the one-point and two-point Green
functions of this operator are calculated in certain limit.  Then
we analysis the divergence behavior of the Green functions, and
find radial/energy-scale relation for this gauge/gravity duality
configuration. Some further results and $\beta$-function of
$\net$ SYM are also achieved in this section. Finally, a brief
summary is in section 5.

\section{Wrapped D5 brane solution and $\net,\;D=4$ SYM}

In this section we review wrapped D5 branes configuration which
is dual to pure $\net$ SYM theory in four dimensions.

The essential idea is to consider the $SO(4)$ gauged supergravity
in $D=7$\cite{SS83}, find its domain-walls solution which wraps on
a 2-sphere, and then lift it up in ten
dimensions\cite{MN01,MN01b}. This $SO(4)$ gauged SUGRA can be
obtained via perform an $S^3$ reduction in type IIB
SUGRA\cite{CLP00}. In order to obtain a solution dual to $\net$
SYM which preserves eight supercharges, one should truncate the
$SO(4)$ gauge group of 7-d SUGRA to its $U(1)$ subgroup. The
Lagrangian for this truncated gauged SUGRA is
\begin{eqnarray}\label{2.1}
{\cal L}_7&=&\sqrt{-{\rm
det}G}\{R-\frac{5}{16}\pa_iy\pa^iy-\pa_ix\pa^ix-\frac{1}{4}
e^{-2x-y/2}F_{ij}F^{ij} \nonumber \\
&&-\frac{1}{12}e^{-y}H_{ijk}H^{ijk}+4e^{y/2}/r_0^2\},
\end{eqnarray}
with
$$F_{ij}=\pa_iA_j-\pa_jA_i, \hspace{0.5in}
  H_{ijk}=\pa_iA^{(2)}_{jk}-\pa_jA^{(2)}_{ik}+\pa_kA^{(2)}_{ji}.$$
Here $i,j,k=0,...6$ are 7-d space-time indices, $r_0$ is constant
with length dimension, $x,\;y$ are scalars fields yielded from
10-d dilation and dimensional reduction, $A_i$ is $U(1)$ gauge
field and $A^{(2)}$ are two-form potential inheriting from type
IIB SUGRA.

The metric ansatz for the domain-wall solution is
\begin{eqnarray}\label{2.2}
ds_7^2=e^{2f(r)}(dx_{1,3}^2+dr^2)+r_0^2e^{2g(r)}d\Omega_2^2
\end{eqnarray}
where $dx_{1,3}^2$ is the Minkowski metric on ${\bf R}_{1,3}$,
$r$ is the transverse coordinate to the domain-wall, and
$d\Omega_2^2=d\theta^2+\sin^2\theta
d\vphi^2\;\;(0\leq\theta\leq\pi,\;0\leq\vphi\leq 2\pi)$ is the
metric of unit 2-sphere. To implement the topological twist that
preserves eight supercharges, we have to identify the $U(1)$
gauge field with the spin-connection on the tangent bundle to the
sphere, i.e., $A=r_0^{-1}\cos\theta d\vphi$. It is also
consistent to set $A^{(2)}=0$ and scalar fields $x,\;y$ are only
dependent on $r$. Then to find the domain-wall of this truncated
gauged SUGRA is just to solve the field equations of scalar
functions $f(r),\;g(r),\;x(r)$ and $y(r)$.

Since all scalar functions are independent of $\theta$ and
$\vphi$, one can further perform an $S^2$ reduction in this 7-d
SUGRA, the result action is
\begin{eqnarray}\label{2.3}
S_5&=&\eta_5\int d^4xdr\;e^{2k}\left\{4\pa_\mu k\pa^\mu k-2\pa_\mu
h\pa^\mu h-\pa_\mu x\pa^\mu x \right.\nonumber \\
&&\left.+4\left(\frac{dk}{dr}\right)^2-2
\left(\frac{dh}{dr}\right)^2-\left(\frac{dx}{dr}\right)^2-V(x,h)
\right\},
\end{eqnarray}
where we have imposed the relation $y=-4f$ \cite{GKM01} such that
\begin{eqnarray}\label{2.4}
&&h=g-f,\hspace{1.1in}k=\frac{3}{2}f+g, \nonumber \\
&&V(x,h)=-r_0^{-2}(4+2e^{-2h}-\frac{1}{2}e^{-4h-2x}),
\end{eqnarray}
$\eta_5$ is normalization constant which in principle has 10-d
origin, and 4-d space-time indices $\mu,\;\nu$ are raised or
lowered by Minkowski metric.

Then the domain-wall solution is conveniently represented by the
variable $z=e^{2h}$
 \beq{2.5}
 e^{2k+x}=ze^{2z},\hspace{0.6in} e^{-2x}=1-\frac{1+ce^{-2z}}{2z},
 \eeq
 where $c$ is integration constant. This solution satisfies the
following first-order Hamilton equations
 \beq{2.6}
  \frac{dk}{dr}=\frac{1}{4}{\cal W},\hspace{0.3in}
  \frac{dh}{dr}=-\frac{1}{4}\frac{\pa{\cal W}}{\pa
  h},\hspace{0.3in}
  \frac{dx}{dr}=-\frac{1}{2}\frac{\pa{\cal W}}{\pa x},
 \eeq
with
 \beq{2.7}
   {\cal W}(x,h)=-r_0^{-1}(4\cosh{x}+e^{-2h-x}).
 \eeq
The following steps are substituting solution~(\ref{2.5}) such
that $f,\;g$ into 7-d metric~(\ref{2.2}), and up-lift it to ten
diemensions\cite{GKM01,CLP00}. It describes the NS5 brane
configuration. So that we can obtain the D5 brane configuration
via performing S-duality\cite{GKM01,DLM02}. The 10-d metric in
string frame is given by
 \beq{2.8}
  ds_{10}^2&=&e^{\omega}\{dx_{1,3}^2+zr_0^2d\Omega_2^2
   +e^{2x}r_0^2dz^2+r_0^2d{\td \theta}^2 \nonumber \\
   &&+\frac{r_0^2}{\Delta}(e^{-x}\cos^2{\td \theta}
   (d{\td \vphi_1}^2+\cos{\theta}d\vphi)^2
   +e^{x}\sin^2{\td \theta}d{\td \vphi_2}^2)\},
 \eeq
with
 \beq{2.9}\Delta=e^x\cos^2{\td \theta}+e^{-x}\sin^2{\td\theta},\eeq
and 10-d dilation given by
 \beq{2.10}e^{2\omega}=e^{2z}(1-\sin^2{\td \theta}
   \frac{1+ce^{-2z}}{2z}).\eeq
In addition, a magnetic R-R 2-form is present. It determines the
parameter $r_0$ in terms of the number of wrapped D5 branes, $N$,
and string parameters
 \beq{2.11} r_0^2=Ng_s\alpha'. \eeq

In order to make the structure of D5 branes clearer, one can
define $H=e^{-2\omega}$ and\cite{DELI02}
 \beq{2.12}\rho=r_0\sin{\td \theta}e^{z},\hspace{0.6in}
  \sigma=r_0\sqrt{z}\cos{\td\theta}e^{z-x}.
  \eeq
Such that we have
 \beq{2.13}
   ds_{10}^2&=&H^{-1/2}(dx_{1,3}^2+zr_0^2d\Omega_2^2)
   +H^{1/2}(d\rho^2+\rho^2d{\td \vphi_2}^2) \nonumber \\
   &&+\frac{H^{1/2}}{z}\{d\sigma^2+\sigma^2(d{\td\vphi}_1
    +\cos{\theta}d\vphi)^2)\}.
 \eeq
It should be noted this solution does not hold maximal
supersymmetry, but only preserves eight independent Killing
spinors\cite{GKM01}. They project in eight space-time directions,
namely ${\td \theta}=\pi/2$ in metric~(\ref{2.8})\cite{DLM02}. So
that we will conveniently set ${\td \theta}=\pi/2$ in our
following calculations.

\section{Fluctuation solution}

In this paper we would like to study 1-point and 2-point Green
functions of the following dimension four operator
 \beq{3.1}{\cal O}(x)=Tr(D_\mu\Phi^\dag D^\mu\Phi
 +2\bar{\Psi}_A\Sla{D}\Psi^A)-\frac{1}{2}Tr(F_{\mu\nu}F^{\mu\nu}),
 \eeq
where $(A_\mu,\;\Psi_A,\;\Phi)$ forms a $\net$ vector
supermultiplet, and $Tr$ taken over $SU(N)$ gauge group. It is
unambiguous that the fluctuation of gauge coupling $1/g_{YM}^2$
couples to this operator. Therefore, in order to derive field
equation of fluctuation field, we should first find which fields
of $h,\;x$ or $k$ determine the gauge coupling of $\net$ SYM.

For achieving this purpose, we have to up-lift metric~(\ref{2.2})
to ten dimensions without considering any special solutions. In
string frame, the 10-d metric and dilation $\omega$ for D5 brane
configuration are given by
 \beq{3.2}ds_{10}^2&=&e^{\omega}[dx_{1,3}^2+dr^2+r_0^2e^{2h}
   d\Omega_2^2]+..., \nonumber \\
  e^{\omega}&=&e^{5f}\Delta, \eeq
where $\Delta$ has been defined in Eq.~(\ref{2.9}). The
Dirac-Born-Infeld (DBI) action for D5 brane is
 \beq{3.3} {\cal L}_{DBI}&=&-\tau_5\int d^6\xi e^{-\omega}
  \sqrt{-{\rm det}(G+2\pi\alpha' F)}+..., \nonumber \\
  \tau_5&=&(2\pi)^{-5}g_s^{-1}\alpha'^{-3}, \eeq
where $G$ and $F$ are pull-back of 10-d metric and gauge fields.
Conveniently, we can parameterize brane world-volume coordinates
by $\xi=\{x^0,...,x^3,\theta,\vphi\}$. Then integrating the
compact part of D5 brane, we will achieve 4-d $\net$ SYM theory
at leading order of $\alpha'$ expansion. The gauge coupling of
$\net$ SYM is given by
 \beq{3.4}\frac{1}{g_{YM}^2}&=&\frac{\tau_5(2\pi)^2\alpha'^2}{2}
 \int_0^{2\pi}d\vphi\int_0^\pi d\theta e^{-3\omega}\sqrt{-{\rm
 det}G}, \nonumber \\
 &=&\frac{N}{4\pi^2}e^{2h}. \eeq
Now let
 \beq{3.5}e^{2h}\;\rightarrow\;e^{2h(r)}+\left(\frac{N}{4\pi^2}
   \right)^{-1}\phi(x,r)=e^{2h(r)}+{\td \phi}(x,r). \eeq
Thus field $\phi(x,r)$ is fluctuation of domain-wall solution,
$z=e^{2h}$, and its boundary value $\phi(x,r_c)$ couples to
operator ${\cal O}(x)$. It is in general consistent to consider
the following ansatz
 \beq{3.6}{\td \phi}(x,r)=s(r)\int\frac{d^4q}{(2\pi)^2}e^{iq\cdot x}
 \eeq
Then inserting Eq.~(\ref{3.5}) into action~(\ref{2.3}) and
expanding it up to quadratic terms of ${\td \phi}$, we get
 \beq{3.7}\delta S_5&=&\eta_5\int d^4x dr\;e^{2k-4h}{\big\{}
 -\frac{1}{2}(\frac{d{\td\phi}}{dr})^2+2\frac{dh}{dr}
 \frac{d{\td\phi}^2}{dr}-M^2{\td\phi}^2{\big\}},
 \eeq
where
 \beq{3.8}M^2=6(\frac{dh}{dr})^2-\frac{2}{r_0^2}e^{-2h}
  +\frac{3}{2r_0^2}e^{-4h-2x}-\frac{q^2}{2}.\eeq
The linear terms of ${\td\phi}$ vanish due to field equation of
$h$, up to a surface term.

The field equation yielded by action~(\ref{3.7}) is
 \beq{3.9} \frac{d^2s}{dr^2}+2\frac{d(k-2h)}{dr}\frac{ds}{dr}
 +\left(q^2+4(\frac{dh}{dr})^2-r_0^{-2}e^{-4h-2x}\right)s=0. \eeq
This equation does not possess analytic solution for complete
expression of background~(\ref{2.5}) and (\ref{2.6}). However,
since we focus on divergent behavior and Green functions, in fact
we only need to know the behavior of $\phi(x,r)$ close to
boundary. From metric~(\ref{2.13}) and (\ref{2.10}) we see that
there are two possible ``boundaries'' for this geometry when ${\td
\theta}=\pi/2$,
 \begin{equation}\label{3.10}\left\{ \begin{array}{ll}
   z\rightarrow\infty &\hspace{1in} {\rm for\;\;all\;\;c} \\
   z=0 & \hspace{1in}{\rm for}\;\; c<-1
   \end{array} \right. \end{equation}
We are not interesting to second case since the condition $c<-1$
does not allow D5 branes existing in fact. Then at near boundary
limit $z\rightarrow\infty$, the domain-wall solution~(\ref{2.5})
and (\ref{2.6}) is simplified,
 \beq{3.11}e^{2x}&=&1,\hspace{1.35in} e^{2k}=ze^{2z}, \nonumber \\
 \frac{dk}{dr}&=&-r_0^{-1}(1+\frac{1}{4z}),\hspace{0.5in}
 \frac{dh}{dr}=-\frac{1}{2r_0z}.
 \eeq
Substituting this background together with $e^{2h}=z$ into field
equation~(\ref{3.9}), we have
 \beq{3.12}\frac{d^2s}{dz^2}+(2-\frac{3}{2z})\frac{ds}{dz}
   +r_0^2q^2s=0. \eeq
This equation has two asymptotical solutions for large $z$:
\begin{equation}\label{3.13}
 \begin{array}{llll}
 a) & s(z)\rightarrow z^{-m_1}e^{a_1z}, &\hspace{0.5in}
   a_1=-1+\sqrt{1-r_0^2q^2},
   & m_1=\frac{3}{2}\frac{1-\sqrt{1-r_0^2q^2}}{\sqrt{1-r_0^2q^2}};
     \\
 b) & s(z)\rightarrow z^{m_2}e^{a_2z}, &\hspace{0.5in}
   a_2=-1-\sqrt{1-r_0^2q^2},
   & m_2=\frac{3}{2}\frac{1+\sqrt{1-r_0^2q^2}}{\sqrt{1-r_0^2q^2}}.
 \end{array}
\end{equation}
Again, similar to AdS$_5$ case, the solution $a)$ is
non-normalizable and $b)$ is normalizable. Since $r_0^2q^2\sim
\alpha'q^2\to 0$ for any low energy field theory, we further have
\beq{3.14} a_1&=&-\frac{1}{2}r_0^2q^2, \hspace{1.05in}
   m_1=\frac{3}{4}r_0^2q^2, \nonumber \\
   a_2&=&-2+\frac{1}{2}r_0^2q^2, \hspace{0.8in}
   m_2=3+\frac{3}{4}r_0^2q^2 \eeq
Imposing the boundary condition $\phi(x,z)=\phi_0(x)=e^{iq\cdot
x}$ at $z=z_0\to \infty$, we find the asymptotical solution of
${\td \phi}$ for large $z$ as follows
 \beq{3.15}{\td\phi}(x,z)\to\frac{\lambda_1z^{-m_1}e^{a_1z}
  +\lambda_2z^{m_2}e^{a_2z}}{\lambda_1z_0^{-m_1}e^{a_1z_0}
  +\lambda_2z_0^{m_2}e^{a_2z_0}}\int\frac{d^4q}{(2\pi)^4}
  e^{iq\cdot x}, \eeq
where $\lambda_1$ and $\lambda_2$ are arbitrary constants.

\section{Green functions and radial/energy-scale relation}

\subsection{Green functions}

In SUGRA picture the one-point function of an operator
corresponds to the first variation of the SUGRA action. In
general, this quantity is expected to vanish due to field
equation. However, the first variation is only required to vanish
up to a total derivation term. Indeed, unlike conformal case,
there is a total derivation term which belongs to linear term of
${\td\phi}$,
 \beq{4.1} S_1=-2\eta_5\int d^4xdr\;\frac{d}{dr}(e^{2k-2h}
  \frac{dh}{dr}{\td\phi}) \eeq
It leads to one-point function of operator ${\cal O}(x)$ as
follows
 \beq{4.2}<{\cal O}(x)>=\frac{\delta S}{\delta \phi_0(x)}
  {\Big |}_{z=z_0,\phi_0=0}=r_0^{-1}\eta_5\frac{e^{2z_0}}{z_0}.
  \eeq
In addition, two-point function is given by
 \beq{4.3}
 &&<{\cal O}(p){\cal O}(q)> \nonumber \\
 &=&\frac{\delta^2 S}
 {\delta\phi_0(p)\delta\phi_0(q)}{\Big |}_{z=z_0,\phi_0=0}
  =2\eta_5\delta^4(p+q)e^{2k-4h}(-\frac{1}{2}s\frac{ds}{dr}
   +2s^2\frac{dh}{dr}){\Big |}_{z=z_0} \nonumber \\
  &=&\frac{\eta_5}{2r_0}\delta^4(p+q)\frac{e^{2z_0}}{z_0}
   \left[-\frac{r_0^2q^2z_0}{2}-\frac{3r_0^2q^2}{4}
   -\frac{\lambda_2}{\lambda_1}(2+r_0^2q^2)
   z_0^{4+\frac{3r_0^2q^2}{2}}e^{-(2+r_0^2q^2)z_0}
  \right] \nonumber \\
  &&+O(r_0^4q^4)+O(e^{-4z_0}).
  \eeq
The result~(\ref{4.3}) is notable. In some terms momenta
$r_0^2q^2\sim\alpha'q^2$ behaviors exponentially, as like as
Veneziano amplitude\cite{Vene74} at the hard scattering limit in
perturbative string theory. It is obvious that this contribution
can not be obtained from perturbative calculation of QFT.
Precisely, it indicates that some non-local interactions are
turned on. In other words, the operator ${\cal O}(x)$ describes a
composite object which will be not pointlike at string scale. In
next subsection we will show that, at leading order of $\alpha'$
expansion, it can be interpreted as non-perturbative instanton
effect.

\subsection{Radial/energy-scale relation}

According to Eqs.~(\ref{2.12}) and (\ref{3.4}), the two-point
function of ${\cal O}(x)$ can be re-expressed by radial parameter
$\rho$ and Yang-Mills coupling $g_{YM}^2$,
 \beq{4.4}<{\cal O}(p){\cal O}(q)>\sim
 -\eta_5\frac{Ng_{YM}^2}{16\pi^2}\frac{\rho_0^2}{r_0}q^2\delta^4(p+q)
 +O(\rho_0^2/\ln{\rho_0^2})+O(g_{YM}^4), \eeq
where $\rho_0\to\infty$ is large radial cut-off. In the above
equation we have used the fact that large radial limit in SUGRA
corresponds to weak 't Hooft coupling limit, $Ng_{YM}^2<<1$, in
SYM. It is same to $AdS_5$/CFT$_4$ case that, at the decoupling
limit $\alpha'\to 0$, we expect a new radial variable
$\nu=\frac{r_0^2}{\rho}$ should be independent of $\alpha'$. Then
we have
 \beq{4.5}<{\cal O}(p){\cal O}(q)>\sim
 -r_0^3\eta_5\frac{Ng_{YM}^2}{16\pi^2}\ep^{-2}q^2\delta^4(p+q)
 +O(1/(\ep^2\ln{\ep^2}))+O(g_{YM}^4), \eeq
where $\ep=\nu_0\to 0$ also is a cut-off. Because
$\eta_5\sim\alpha'^{-3/2}$, this result is independent of
$\alpha'$, same as the result of QFT. Furthermore, the
perturbative calculation of QFT yields divergent structure of
two-point function of ${\cal O}$ as follows
 \beq{4.6} <{\cal O}(p){\cal O}(q)>\sim g_{YM}^2
  \left(\alpha_0\Lambda^4
  +\alpha_1\Lambda^2q^2+O(\ln{\frac{\Lambda^2}{q^2}})\right)
  \delta^4(p+q). \eeq
Comparing $q^2$ terms in Eqs.~(\ref{4.5}) and (\ref{4.6}), we
obtain $\ep\sim \Lambda^{-1}$. Here we haven taken cut-off both
for radial in gravity and for energy in QFT. If let them flow
away from cut-off point, we have
 \beq{4.7} \nu\sim\mu^{-1}\;\Longrightarrow\;
   z&=&\ln{\mu/M}+C_0, \eeq
where $M$ is definite mass parameter and $C_0$ is an unimportant
constant. This result precisely agrees with result in
Ref.\cite{DLM02}.

The $\beta$-function of $\net$ SYM can be directly obtained from
Eqs.~(\ref{3.4}) and (\ref{4.7}),
 \beq{4.8}\beta(g_{YM})=-\frac{1}{g_{YM}^3}\frac{d}{d\ln{(\mu/M)}}
  \frac{1}{g_{YM}^2}=-\frac{N}{8\pi^2}g_{YM}^3. \eeq
It precisely agrees with the result from perturbative calculation
of QFT.

Now let us consider non-perturbative effect in two-point
function~(\ref{4.3}). According to the identity
 \beq{4.9}\lim_{z_0\to\infty}z_0^3e^{-(2+r_0^2q^2)z_0}
   =\frac{6}{(2+r_0^2q^2)^3}e^{-(2+r_0^2q^2)z_0}, \eeq
at low energy limit $r_0^2q^2\to 0$ the $q^2$ terms of two-point
can be written as
  \beq{4.10} <{\cal O}(p){\cal O}(q)>\sim
 -r_0^3\eta_5\frac{Ng_{YM}^2}{16\pi^2}(\ep^{-2}
 -6r_0^{-2}\frac{\lambda_2}{\lambda_1})q^2\delta^4(p+q)
 \eeq
Comparing with the result of QFT, we achieve
 \beq{4.11} &&\ep^{-2}-6r_0^{-2}\frac{\lambda_2}{\lambda_1}
  \sim \Lambda^2 \nonumber \\
  &\Longrightarrow& z=\frac{1}{2}\ln{(\frac{\mu^2}{M^2}
  +6\frac{\lambda_2}{\lambda_1})}+C_0.
 \eeq
This radial/energy-scale relation leads $\beta$-function of
$\net$ SYM as follows
 \beq{4.12}\beta(g_{YM})=-\frac{N}{8\pi^2}g_{YM}^3
   \left(1+b {\rm exp}\{-\frac{8\pi^2}{Ng_{YM}^2}\}
   +O({\rm exp}\{-\frac{16\pi^2}{Ng_{YM}^2}\})\right),
 \eeq
where $b$ is an unknown constant. The extra term in the above
expression is just non-perturbative contribution from instantons
with fractional charge $\frac{k}{N}$ where $k$ is a positive
integer, as like as one has shown in pure ${\cal N}=1$
SYM\cite{DHKM99}. It would be interesting subject to investigate
whether origin of these fractional instantons are related to
fractional D3 branes.

\subsection{Further discussions}

The first notable fact is one-point function of ${\cal O}$. The
non-zero value of this one-point function implies that the bosons
in $\net$ supermultiplet have condensation. The one-point
function in Eq.~(\ref{4.2}) can be rewritten as
 \beq{4.13} <{\cal O}(x)>=r_0^3\eta_5\frac{Ng_{YM}^2}{4\pi^2}
  \ep^{-4}{\rm exp}\{-\frac{8\pi^2}{Ng_{YM}^2}\}. \eeq
This result matches with QFT calculation $<{\cal
O}>\sim\Lambda^4$. However, the exponential factor indicates that
this condensation entirely is non-perturbative effect induced by
fractional instantons.

It is well-known that the QFT description is reliable at weak
't~Hooft coupling, $Ng_{YM}^2<<1$. However, the SUGRA description
is reliable when radial is much large than string scale, i.e.,
$r_0^2/\alpha'\sim Ng_{YM}^2>>1$. So that SUGRA description on
wrapped D5 brane configuration at weak coupling is dual to a
strong coupled SYM in four dimensions, and vice versa. Thus it is
not surprised why non-perturbative effects appear in Green
functions. It is also same to our usual understanding on
gauge/gravity duality.

In previous subsection the $\beta$-function for $\net$ SYM is
obtained directly when radial/energy-scale relation is imposed,
even without considering any quantum corrections in SYM. It
implies that some quantum effects have been included when we
obtain $\net$ SYM from wrapped D5 brane configuration. The
essential reason is that the coupling of SYM obtained in this way
is no longer constant, but is radial-dependent. Extremely, in
Refs.\cite{DLM02,Im02} the authors showed that not only one-loop
effects, but all of quantum effects is encoded in SYM. It is not
surprised, since when we obtain a four-dimensional gauge theory
via compactifing a six-dimensional theory in non-flat background,
some higher dimensional gravity effects must be involved into
four-dimensional theory. However, a confusion question appears:
if we treat $\net$ SYM as low energy theory of open string theory
at defined limit, but without considering gauge/gravity
correspondence, whether and/or how should we consider quantum
corrections in this low energy theory? Whether is it double
counting if we consider quantum correction? This confusion can be
resolved when we are aware that the gauge theory is defined at the
fixed boundary of space-time geometry. In this sense all couplings
of SYM are still kept as constants at classical level, and the
quantum correction can be consistent included via perturbative
calculation of QFT but without any double counting. However, when
we impose radial/energy-scale relation and let the definition of
gauge theory flow away from boundary, in prior we have included
the effects of gauge/gravity duality. Thus quantum effects are
naturally yielded even without QFT calculation.

\section{Conclusions}

We have evaluated one-point and two-point Green functions of
operator~(\ref{3.1}) in $\net$ SYM according to gauge/gravity
correspondence. The dual SUGRA describes a configuration that D5
branes wrap on supersymmetric 2-cycle. We analyzed divergence
behavior of these Green functions and compared them with one
obtained from perturbative calculation of QFT. Then we achieved
radial/energy-scale relation for this ${\cal N}=2$ gauge/gravity
duality configuration. In terms of this relation, we obtained the
$\beta$-function of $\net$ SYM with fractional instanton
contribution. All results match with one obtained from field
theory. Perturbatively, our result also agrees with earlier result
by authors of Ref.\cite{DLM02}. We also showed that the bosons in
$\net$ supermultiplet have condensation due to instanton effects,
which in general can not be observed in perturbative QFT.

Our investigation provides a principle method to study
radial/energy-scale relation for any configuration on
gauge/gravity correspondence. However, we will meet a technical
difficulty when we generalize it to ${\cal N}=1$ case and want to
re-examine ${\cal N}=1$ radial/energy-scale relation in
Ref.\cite{DLM02,Im02}. That the gauge coupling in $\net$ case is
determined by a single background field (see Eq.~(\ref{3.4})),
but in ${\cal N}=1$ it is expressed by nonlinear combination of
several background fields\cite{DLM02,Im02}. Thus it is difficult
to derive field equation of fluctuation of gauge coupling. To
overcome this difficulty will be important to study
radial/energy-scale relation for any gauge/gravity correspondence.

\section*{Acknowledgments}

We thank Prof. E. Imeroni, J.-X. Lu and Ch.J. Zhu for useful
discussions and comments.


\begin{thebibliography}{99}
\bibitem{Mald98}J. Maldacena, {\sl The large N limit of superconformal
field theories and supergravity}, Adv. Theor. Math. Phys. {\bf 2}
(1998) 231.
\bibitem{Orbifold}S. Kachru and E. Silverstein, {\sl 4d Conformal
Field Theory and Strings on Orbifold}, Phys. Rev. Lett. {\bf 80}
(1998) 4855; A. Lawrence, N. Nekrasov and C. Vafa, {\sl On
Conformal Field Theory in Four Dimensions}, Nucl. Phys. {\bf
B533} (1998) 199.
\bibitem{Conifold}I.R. Klebanov and E. Witten, {\sl Superconformal
Field Theory on Threebranes at A Calabi-Yau Singularity}, Nucl.
Phys. {\bf B536} (1998) 199; K. Oh and R. Tatar, {\sl
Renormalization Group Flows on D3 Branes at an Orbifolded
Conifold}, JHEP {\bf 05} (2000) 030; I.R. Klebanov and M.J.
Strassler, {\sl Supergravity and a Confining Gauge Theory:
Duality Cascades and $\chi$SB Resolution of Naked Singularities},
JHEP {\bf 08} (2000) 052.
\bibitem{Fbrane}D. Diaconescu, M.R. Douglas and J. Gomis, {\sl
Fractional Branes and Wrapped Branes}, JHEP {\bf 02} (1998) 013;
N.A. Nekrasov, {\sl Gravity Duals of Fractional Branes and
Logarithmic RG Flow}, Nucl. Phys. {\bf 574} (2000) 263.
\bibitem{BHreview}M. Bertolini, P. Di Vecchia and R. Marotta, {\sl
$\net$ Four-Dimensional Gauge Theories From Fractional Branes} ,
hep-th/0112195; C.P. Herzog, I.R. Klebanov and P. Ouyang, {\sl
D-branes on The Conifold and ${\cal N}=1$ Gauge Gravity
Dualities}, hep-th/0205100.
\bibitem{BVS96}M. Bershazki, C. Vafa and V. Sadov, {\sl D-branes
and Topological Field Theories}, Nucl. Phys. {\bf B463} (1996)
420.
\bibitem{MN01}J. Maldacena and C. Nunez, {Towards the Large N
Limit of ${\cal N}=1$ Super Yang-Mills}, Phys. Rev. Lett, {\bf
86} (2001) 588.
\bibitem{GKM01}J.P. Gauntlett, N. Kim, D. Martelli and D. Waldram,
{\sl Wrapped Fivebranes and The $\net$ Super Yang-Mills Theory},
Phys. Rev. {\bf D64} (2001) 106008.
\bibitem{BCZ01}F. Bigazzi, A.L. Cotrone and A. Zaffaroni, {\sl
$\net$ Gauge Theories from Wrapped Five-Branes}, Phys. Lett. {\bf
B519} (2001) 269; R. Apreda, F. Bigazzi, A. L. Cotrone, M.
Petrini and A. Zaffaroni, {\sl Some Comments on N=1 Gauge
Theories from Wrapped Branes}, Phys. Lett. {\bf B536} (2002) 161.
\bibitem{review2}Y. Kinar, A. Loewy, E. Schreiber, J.
Sonnenschein and S. Yankielowicz, JHEP {\bf 03} (2001) 013; B.S.
Acharya, J.P. Gauntlett and N. Kim, Phys. Rev. {\bf D63} (2001)
106003; H. Nieder and Y. Oz, JHEP {\bf 03} (2001) 008; J. Gomis,
Nucl. Phys. {\bf B606} (2001) 3; G. Papadopoulos and A.A.
Tseytlin, Class. Quant. Grav. {\bf 18} (2001) 1333; Jerome P.
Gauntlett, Nakwoo Kim, Daniel Waldram, Phys. Rev. {\bf D63}
(2001) 126001; C. Nunez, I.Y. Park, M. Schvellinger and T.A.
Tran, JHEP {\bf 04} (2001) 025; M. Schvellinger and T. A. Tran,
JHEP {\bf 06} (2001) 025; J. Gomis and J.G. Russo, JHEP {\bf 10}
(2001) 028; R. Hernandez, Phys. Lett. {\bf B521} (2001) 371; J.
Gomis and T. Mateos, Phys. Lett. {\bf B524} (2002) 170; J. Gomis,
Nucl. Phys. {\bf B624} (2002) 181; N. Evans, M. Petrini, A.
Zaffaroni, JHEP {\bf 06} (2002) 004; U. Gursoy, C. Nunez and M.
Schvellinger, JHEP {\bf 06} (2002) 015.
\bibitem{BVV00}J. de Boer, E. Verlinde and H. Verlinde, {\sl On
the Holographic Renormalization Group}, JHEP {\bf 08} (2000) 003.
\bibitem{PP98}A.W. Peet and J. Polchinski, {\sl UV/IR relations in
AdS Dynamics}, Phys. Rev. {\bf D59} (1999) 065011.
\bibitem{DLM02}P. Di Vecchia, A. Lerda and P. Merlatti, {\sl ${\cal
N}=1$ and $\net$ Super Yang-Mills Theories from Wrapped Branes},
hep-th/0205204.
\bibitem{Seiberg88}N. Seiberg, Phys. Lett. {\bf B206} (1988) 75;
N. Seiberg and E. Witten, Nucl. Phys. {\bf B426} (1994) 19; {\sl
ibid}, {\bf B431} (1994) 484.
\bibitem{SS83}A. Salam and E. Sezgin, {\sl SO(4) Gauging of $\net$
Supergravity in Seven Dimensions}, Phys. Lett. {\bf B126} (1983)
295.
\bibitem{MN01b}J. Maldacena and C. Nunez, {\sl Supergravity
Description of Field Theories on Curved Manifold and A No-Go
Theorem}, Int. J. Mod. Phys. {\bf A16} (2001) 822.
\bibitem{CLP00}M. Cvetic, H. Lu and C.N. Pope, {\sl Consistent
Kaluza-Klein Sphere Reduction}, Phys. Rev. {\bf D62} (2000)
064028.
\bibitem{DELI02}P. Di Vecchia, H. Enger, E. Lozano-Tellechea and
E. Imeroni, {\sl Gauge theories from wrapped and fractional
branes}, Nucl. Phys. {\bf B631} (2002) 95.
\bibitem{Vene74}G. Veneziano, {\sl An Introduction to Dual Models
of Strong Interactions and Their Physical Motivations}, Phy. Rep.
{\bf 9} (1974) 199.
\bibitem{DHKM99}N.M. Davies, T.J. Hollowood, V.V. Khoze and M.P.
Mattis, {\sl Gluino Condensate and Magnetic Monopoles in
Supersymmetric Gluodynamics}, Nucl. Phys. {\bf B559} (199) 123.
\bibitem{Im02}E. Imeroni, {\sl On the ${\cal N}=1$
$\beta$-function from the Conifold}, hep-th/0205216.
\end{thebibliography}
\end{document}